\documentclass[aps,prl,reprint,groupedaddress,showpacs]{revtex4-1}
\usepackage{amsmath,amssymb, epsfig}

\begin{document}


\title{A Classical Nernst Engine}



\author{Julian Stark,\textsuperscript{1}
Kay Brandner,\textsuperscript{1} Keiji Saito,\textsuperscript{2} and Udo Seifert\textsuperscript{1}}

\affiliation{\textsuperscript{{{\rm 1}}}II. Institut f\"ur Theoretische Physik, Universit\"at Stuttgart, 70550 Stuttgart, Germany\\
\textsuperscript{{{\rm 2}}}Department of Physics, Keio University, 3-14-1 Hiyoshi, Kohoku-ku, Yokohama, Japan 223-8522}

\author{}
\affiliation{}


\begin{abstract}
We introduce a simple model for an engine based on the Nernst effect. 
In the presence of a magnetic field, a vertical heat current can
drive a horizontal particle current against a chemical potential.
For a microscopic model invoking classical particle trajectories
subject to the Lorentz force, we prove a universal bound
$3-2\sqrt{2}\simeq 0.172$ for the ratio between maximum efficiency 
and Carnot efficiency. This bound, as the slightly lower one $1/6$ for 
efficiency at maximum power, can indeed be saturated for large 
magnetic field and small fugacity irrespective of the aspect ratio.
\end{abstract}


\maketitle

\newcommand{\Jout}{J^{\varrho+}_i}
\newcommand{\Jin}{J^{\varrho -}_i}
\newcommand{\ar}{\mathcal{A}}
\newcommand{\vectt}[2]{\left( \! \begin{array}{c}
#1\\ #2 \end{array} \! \right)}
\newcommand{\vecttt}[3]{\left[ \! \begin{array}{c}
#1 \\ #2\\ #3 \end{array}\! \right]}
\newcommand{\matt}[4]{\left( \! \begin{array}{cc}
#1 & #2\\ #3 & #4 \end{array} \! \right)}

\emph{Introduction}.--
The Nernst effect describes the emergence of an electrical voltage 
perpendicular to a heat current transversing an isotropic conductor 
in the presence of a constant magnetic field \cite{Goldsmid2009}.
However, while Seebeck-based devices, for which the heat and the 
particle current are coupled without a magnetic field, have been the
subject of intensive research efforts during the last decades
\cite{Dresselhaus2007, Snyder2008, Bell2008, Vineis2010}, only a few 
attempts were made to utilize the Nernst effect for power generation
\cite{Elliott1959, Wright1962, Norwood1963, Harman1963}.
This lack of interest may have been caused by the uncompetitive net 
efficiency of Nernst-based devices, which is inevitably suppressed
by the energetic cost of the strong magnetic fields they require.
New discoveries in the phenomenological theory of thermoelectric 
effects as well as recent experiments showing the accessibility of 
magnetic field effects in nanostructures even at low and moderate field 
strengths \cite{Pogosov2002, Maximov2004, Goswami2011, Matthews},
however, cast new light on the topic of Nernst engines.

Benenti and co-workers showed by a quite general analysis within
the framework of linear irreversible thermodynamics that breaking
the microscopic time-reversal symmetry by a magnetic field could,
in principle, increase thermoelectric efficiency such that
even devices operating reversibly at finite power seem
to be achievable \cite{Benenti2011}.
Such an intriguing suggestion asks for a better understanding of 
coupled heat and particle transport in magnetic fields. 
First progress in this direction was recently achieved
within the paradigmatic class of multi-terminal models, for which
it turned  out that current conservation implies much stronger
bounds on the efficiency than the standard rules of linear 
irreversible thermodynamics \cite{Brandner2013, Brandner2013a}.
For the minimal case of three terminals, these bounds were 
even shown to be tight \cite{Balachandran2013}.
Since these models were based on general particle transmission
probabilities without reference to any specific microscopic dynamics,
they leave the necessary conditions for saturating these bounds open.

Simple mechanical models have led to remarkable insight into the
microscopic mechanisms underlying heat and matter transport
\cite{Mejia-Monasterio2001, Li2003},
especially in the context of thermoelectric efficiency
\cite{Casati2008, Horvat2009, Saito2010}.
So far, the effect of a magnetic field breaking time-reversal 
symmetry on thermoelectric efficiency has not yet been addressed
using such models.
Nernst engines are ideal candidates to investigate the influence of 
broken time-reversal symmetry.
We therefore propose a minimalistic, classical model for such an 
engine, which provides physical insight on the level of single 
particle trajectories.
\begin{figure}
\epsfig{file=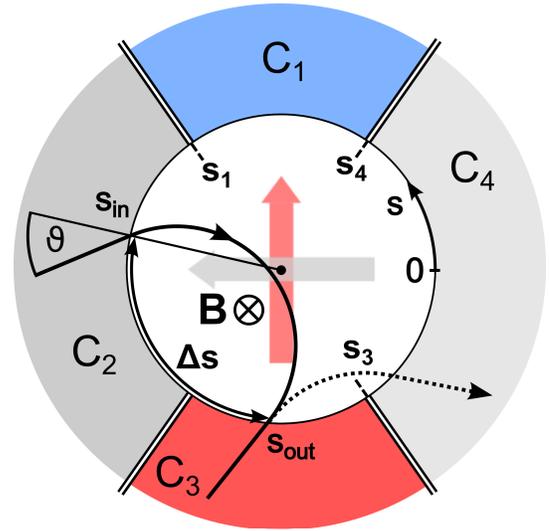, scale=0.40}
\caption{Scheme of the classical Nernst engine. The vertical heat current
(red arrow) between reservoir $C_3$ and $C_1$ with $T_3>T_1$ drives a 
horizontal particle current (grey arrow)
from reservoir $C_4$ to $C_2$ with $\mu_2>\mu_4$.  
The bold black line denotes a classical trajectory leaving reservoir
$C_2$ at $s_{{{\rm in}}}$ with an angle $\vartheta < 0$ and entering $C_3$ at $s_{{{\rm out}}}$
where $0\leq s\leq 2\pi R$ parametrizes the circumference. 
The dotted line shows the corresponding time-reversed trajectory for 
an entry at $s_{{{\rm out}}}$. 
For further symbols, see main text.
\label{Fig_Setup}}
\end{figure}

\emph{System}-- Our classical system for studying transport is inspired 
by the Landauer-B\"uttiker approach, which has proven extremely useful 
in the quantum realm \cite{Buttiker1985, Buttiker1986b}.
As shown in Fig. \ref{Fig_Setup}, we consider a 
two-dimensional, circular and potential-free scattering region of radius
$R$ perpendicularly penetrated by a homogeneous magnetic field 
$\mathbf{B}$ of strength $B\equiv |\mathbf{B}|$, surrounded by four 
thermochemical reservoirs $C_i$.
Due to the Lorentz force, a particle with energy $E$ injected 
from reservoir $C_i$ at $s_{{{\rm in}}}$ with an angle $\vartheta \in \Delta$,
where $\Delta\equiv [-\pi/2,\pi/2]$, moves on a circle of radius 
\begin{equation}
R_c=\sqrt{2mE}c/(|q|B)\equiv \nu(E)R 
\end{equation}
inside the scattering region. Here, $c$ denotes the speed of light,
$q<0$ the charge of the particle and $m$ its mass.
This particle hits the boundary at a position $s_{{{\rm out}}}\equiv
s_{{{\rm in}}}+\Delta s$.
A simple geometrical analysis shows for the distance $\Delta s$ measured along the boundary
\begin{equation}\label{Delta_s}
\Delta s = 2R 
\begin{cases}
g(\vartheta, \nu) + \pi & {{\rm for}} \quad   \nu>1, \; 
\sin \vartheta < - 1/\nu \\
g(\vartheta, \nu)       & {{\rm else}}         
\end{cases}
\end{equation}
with 
\begin{equation}
g(\vartheta,\nu)\equiv {{\rm arccot}}\left[(1+\nu\sin\vartheta)/
(\nu\cos\vartheta)\right].
\end{equation}
Note that from (\ref{Delta_s}) onwards, we suppress in $\nu$ the 
dependence on the energy $E$ to simplify the 
notation.

The fluxes entering and leaving the system through the reservoirs
can be determined as follows. 
Any particle that reaches the circular boundary from one of the 
reservoirs is assumed to enter the scattering region. 
A Maxwell-Boltzmann statistics in reservoirs modeled as ideal gases
with inverse temperature $\beta_i$ and chemical potential $\mu_i$
then implies a total particle current
\begin{equation}\label{Jout}
\Jout \! \equiv
\int_{l_i} \! \! ds \! 
\int_0^{\infty} \! \! \! dE \! 
\int_{\Delta} \! \! d\vartheta \; 
u_i(E)\cos\vartheta = 
\frac{\sqrt{2\pi m}l_i}{\beta_i^{3/2}} e^{\beta_i\mu_i}
\end{equation}
flowing from the reservoir $C_i$ with boundary length $l_i$ into 
the system, where $u_i(E)\equiv \sqrt{2mE}\exp [-\beta_i(E-\mu_i)]$ \cite{Saito2010}.
Likewise, assuming that each particle hitting the boundary from
inside the scattering region is absorbed in the adjacent reservoir,
the steady-state current flowing into $C_i$ reads
\begin{equation}\label{Jin}
\Jin \! \equiv 
\sum_j \int_{l_j} \! \! ds \!
\int_{0}^{\infty} \! \! \! dE \!
\int_{\Delta}\! \!   d\vartheta \; 
u_j(E)\cos \vartheta
\; \tau_i(E,s,\vartheta).
\end{equation}
Note that we set Planck's constant as well as Boltzmann's constant equal to $1$ throughout this letter.
In (\ref{Jin}), we have introduced the conditional probability 
$\tau_i(E,s,\vartheta)$ for a particle of energy $E$  
entering at position $s$ with an angle $\vartheta$ to reach
the boundary of the reservoir $C_i$ after passing 
through the scattering region.
Since we assume purely Hamiltonian
dynamics, this probability can either be $1$ or $0$. 
In order to derive a concise expression for the net particle currents
$J^{\varrho}_i \equiv \Jout-\Jin$, 
we define the transmission coefficients
\begin{equation}\label{Transcoeff}
\mathcal{T}_{ji}(E) \equiv \int_{l_i} ds
\int_{\Delta} d\vartheta \; \tau_j(E,s, \vartheta)
\cos \vartheta.
\end{equation}
As our first main result, we can show that Liouville's theorem implies the sum rules \cite{SM}
\begin{equation}\label{Sum_rules}
\sum_i \mathcal{T}_{ji}(E) = 2l_j
\quad {{\rm and}} \quad
\sum_j \mathcal{T}_{ji}(E) = 2l_i.
\end{equation}
By combining (\ref{Jout}), (\ref{Jin}) and (\ref{Sum_rules}), we 
finally arrive at 
\begin{equation}\label{Jrho}
J^{\varrho}_i = \sum_j \int_{0}^{\infty} dE \; \mathcal{T}_{ij}(E)
\left(u_i(E)- u_j(E) \right).
\end{equation}
An analogous calculation yields the net heat flux leaving reservoir
$C_i$
\begin{equation}\label{Jq}
J^{q}_i = \sum_j \int_{0}^{\infty} dE \; \mathcal{T}_{ij}(E)
(E-\mu_i)\left(u_i(E)- u_j(E) \right).
\end{equation} 

\emph{Nernst Engine}.-- For a Nernst engine, we have to impose the 
boundary conditions 
\begin{equation}\label{BoundaryCond}
J^{\varrho}_1=J^{\varrho}_3=0
\quad {{\rm and}} \quad 
J_2^q=J_4^q=0,
\end{equation}
which ensure that the particle current occurs only horizontally
and heat flows only vertically in the set-up of Fig. \ref{Fig_Setup}. 
From here on, we focus on the linear response regime.
We choose the reference values, $\mu\equiv \mu_2$ and
$T\equiv T_1$ and define $\Delta \mu_i\equiv \mu_i -\mu$ 
and $\Delta T_i \equiv T_i-T$. 
Linearizing the currents (\ref{Jrho}) and (\ref{Jq}) 
with respect to $\Delta T_i$ and $\Delta\mu_i$ yields six 
phenomenological relations
\begin{equation}\label{PhenEqFull}
J^{\kappa}_i = \sum_{j \nu}
L^{\kappa \nu}_{ij}\mathcal{F}^{\nu}_j 
\quad {{\rm with }} \quad 
\kappa, \nu = \varrho, q. 
\end{equation}
Here, we have introduced the affinities $\mathcal{F}^{\varrho}_i 
\equiv\Delta\mu_i/T$ and  $\mathcal{F}^q_i\equiv \Delta T_i/T^2$, and
the Onsager coefficients 
\begin{multline}\label{PrimOC}
\matt{L_{ij}^{\varrho \varrho}}{L_{ij}^{\varrho q}}{L_{ij}^{q\varrho}}
{L_{ij}^{qq}}
\equiv \int_0^{\infty} \! \! \! dE \; u(E)
\matt{1}{E-\mu}{E-\mu}{(E-\mu)^2}\\
\times \left(2l_i  \delta_{ij} 
- \mathcal{T}_{ij}(E)\right)
\end{multline}
with $u(E)\equiv \sqrt{2mE}\exp[-\beta(E-\mu)]$.
Using the constraints (\ref{BoundaryCond}) to eliminate $\mathcal{F}^{
\varrho}_1, \mathcal{F}^{\varrho}_3, \mathcal{F}^q_2$ and $\mathcal{
F}^q_4$ in (\ref{PhenEqFull}) and defining the current vector $\mathbf{
J}\equiv (J^{\varrho}_4, J^q_3)^t$ and the affinity $\boldsymbol{
\mathcal{F}} \equiv (\mathcal{F}^{\varrho}_4, \mathcal{F}^q_3)^t$ 
leaves us with
\begin{equation}\label{RedPhenEq}
\mathbf{J}= \mathbb{L}\boldsymbol{\mathcal{F}},
\quad {{\rm where}} \quad
\mathbb{L}\equiv
\matt{L_{\varrho\varrho}}{L_{\varrho q}}{L_{q\varrho}}{L_{qq}}
\end{equation}
is a matrix of effective Onsager coefficients. 

The role of geometry will be studied by introducing the aspect ratio
$\ar\equiv l_2/l_1$. For the choice $l_1=l_3=\pi R/ (1+\ar)$ and 
$l_2=l_4= \pi R\ar/(1+\ar)$, the resulting mirror symmetry implies 
\begin{equation}\label{Symmetry_OC}
L_{\varrho q}=-L_{q \varrho}.
\end{equation}

For a proper heat engine, we put $\Delta\mu_4<0$ and
$\Delta T_3>0$. 
The generated output power and efficiency then become $P=-\Delta \mu_4 
J^{\varrho}_4$ and $\eta=P/J^q_3$ \cite{Seifert2012}. 
Maximizing $\eta$ with respect to $\mathcal{F}^{\varrho}_4$ under
the condition $P\geq 0$ yields
\begin{equation}\label{MaxEffZT}
\eta_{{{\rm max}}}= \eta_C
\frac{1-\sqrt{1-\mathcal{Z}T}}{1+\sqrt{1-\mathcal{Z}T}}
\quad {{\rm with}} \quad
\mathcal{Z}T\equiv \frac{L_{\varrho q}^2}
{{{\rm Det}}\; \mathbb{L}},
\end{equation}
where $\eta_C\equiv 1- T_1/T_3 \approx T\mathcal{F}^q_3$ denotes
the Carnot efficiency. Obviously, like for conventional thermoelectric
devices \cite{Benenti2011}, the maximum efficiency depends only on 
a single dimensionless quantity, the thermomagnetic figure of merit
$\mathcal{Z}T$.
In the literature \cite{Goldsmid2009}, this parameter is usually 
given in the form $\mathcal{Z}T= (N\! B)^2\sigma T/\kappa$,  where 
$N \! B$ is the thermomagnetic power, $\sigma$ the electric and $\kappa$
the thermal conductivity. 
However, this definition coincides with the one 
given in (\ref{MaxEffZT}), if the transport coefficients $N \! B, 
\sigma, \kappa$ are identified correctly with the effective
Onsager coefficients
\footnote{The standard analysis \cite{Callen1985} gives $\sigma=q^2
L_{\varrho\varrho}/T$ and $\kappa= {{\rm Det}} \; \mathbb{L}/(T^2 L_{
\varrho \varrho})$, where $q$ is the charge of the particles. 
The thermomagnetic power is defined as the ratio $N\! B = V/\Delta T$ 
of the transverse voltage emerging due to a longitudinal temperature 
gradient $\Delta T$ if the transverse electrical current is held at $0$
\cite{Goldsmid2009}.
Putting $J_4^{\varrho}=0$ in (\ref{RedPhenEq}) and solving for
$V= - \Delta \mu_4 /q$ gives $N\! B= V/ \Delta T_3 = L_{\varrho q}/
(TqL_{\varrho\varrho})$.}.
In contrast to the naive expectation, $N\! B$ is negative in our model,
i.e., the net particle current flows from the right to the left in 
Fig. \ref{Fig_Setup}, although particles from the hot reservoir are
deflected in the opposite direction. 
This feature is ultimately a consequence of the boundary conditions
(\ref{BoundaryCond}). 

Two bounds successively constrain the parameter $\mathcal{Z}T$.
First, since the second law requires the rate of entropy production 
$\dot{S}=\boldsymbol{\mathcal{F}}^t\mathbf{J}=\boldsymbol{\mathcal{F}}^t
\mathbb{L}\boldsymbol{\mathcal{F}}$ to be non-negative, the matrix 
$\mathbb{L}$ must be positive semi-definite. 
Due to the symmetry (\ref{Symmetry_OC}), this condition reduces to 
$L_{\varrho\varrho}, L_{qq}\geq0$.
By recalling (\ref{MaxEffZT}) one has \cite{Harman1963}
\begin{equation}\label{Second_LawZT}
0\leq \mathcal{Z}T \leq 1.
\end{equation}
Second, by techniques similar to the ones used in \cite{Brandner2013a},
we can show that the Hermitian matrix 
\begin{equation}
\mathbb{K}\equiv \mathbb{L}+ \mathbb{L}^t 
+i(\mathbb{L}-\mathbb{L}^t)
\end{equation}
has to be positive semi-definite as a consequence of the sum rules 
(\ref{Sum_rules}) \cite{SM}. This constraint can be expressed as 
\begin{equation}\label{New_Bound_OC}
({{\rm Det}} \; \mathbb{K})/4 =  L_{\varrho\varrho} L_{qq}
- L_{\varrho q}^2 \geq 0,
\end{equation}
leading to
\begin{equation}\label{New_BoundZT}
0 \leq \mathcal{Z}T \leq 1/2.
\end{equation}
Obviously, the constraint (\ref{New_BoundZT}), which ultimately 
relies on Liouville's theorem, is stronger than (\ref{Second_LawZT}). 
In particular, while the second law, in principle, allows the maximum
efficiency to approach $\eta_C$ in the limit $\mathcal{Z}T\rightarrow 1$,
the bound (\ref{New_BoundZT}) implies the significantly lower limit
\begin{equation}\label{New_BoundMaxEff}
\eta_{{{\rm max}}}\leq (3-2\sqrt{2})\eta_{{{\rm C}}} \simeq
0.172 \eta_C.
\end{equation}
This universal bound on the efficiency of a classical Nernst engine
is our second main result. It arises from the four-terminal set-up and 
the symmetry (\ref{Symmetry_OC}) but is independent of further details
of the geometry and the strength of the magnetic field. 
In the derivation of this bound, we have nowhere used that the 
trajectories are circular in the scattering region.
Hence, it would also apply if an additional potential acted on the 
particles.

Quite generally, the existence of a bound provokes the question whether it 
can be saturated in any given microscopic model.
For addressing this issue within our model, we need to determine the 
Onsager matrix $\mathbb{L}$ and hence the transmission coefficients 
$\mathcal{T}_{ji}(E)$ explicitly.

\emph{Strong field regime}.-- 
Relation (\ref{Delta_s}) allows to determine the 
$\mathcal{T}_{ji}(E)$ for any $E\geq 0$. 
However, the resulting expressions are quite involved \cite{SM}.
We therefore begin with analyzing the limiting case $\nu\ll 1$.
First, expanding (\ref{Delta_s}) in $\nu$ yields  
\begin{equation}\label{Delta_s_Approx}
\Delta s = 2R \nu \cos\vartheta + \mathcal{O}\left(\nu^2 \right).
\end{equation}
Second, since this quantity is bounded from above by $\Delta s^{\ast}= 
2R\nu \ll R$, we can consistently
assume $\Delta s^{\ast} < \min \{l_1, l_2 \}$,
i.e., particles emitted from the reservoir $C_i$ can either pass to the
adjacent reservoir $C_{i+1}$ or return to $C_i$.
Consequently, we have $\mathcal{T}_{ji}(E)=0$ for $j\neq i,i+1$. 
Moreover, the sum rules (\ref{Sum_rules}) require $\mathcal{T}_{ii}(E)
= 2l_i - \mathcal{T}_{i+1i}(E)$. 
Hence, we are left with calculating $\mathcal{T}_{i+1i}(E)$. 
For this purpose, we recall Fig. \ref{Fig_Setup} and recognize that a
particle injected from the reservoir $C_i$ at a certain position $s_{{\rm in}}$ 
must leap over the distance $\Delta s \geq s_i-s_{{\rm in}}$
to reach reservoir $C_{i+1}$, where $s_i$ marks the contact point
of the reservoirs $C_i$ and $C_{i+1}$. 
By virtue of (\ref{Delta_s_Approx}), this transmission condition 
can be rewritten as 
$\vartheta_- < \vartheta < \vartheta_+$ with
$\vartheta_{\pm} \equiv\pm \arccos \left[ (s_i - s_{{{\rm in}}})/(2R\nu)\right]$.
Finally, by using the definition (\ref{Transcoeff}), we get
\begin{equation}
\mathcal{T}_{i+1i}(E)= \int_{s_i-\Delta s^{\ast}}^{s_i}  
\! \! \! d s_{{{\rm in}}} \int_{\vartheta_-}^{\vartheta_+}
\! \! \!  d \vartheta \; \cos \vartheta = \pi R \nu(E).
\end{equation}
Using the complete set of transmission coefficients 
$\mathcal{T}_{ji}(E)$ to calculate the primary Onsager
coefficients (\ref{PrimOC}) and taking into account 
the auxiliary conditions (\ref{BoundaryCond}) yields, as our third 
main result, the effective Onsager matrix 
\begin{equation}\label{ApproxOM}
\mathbb{L} = 
\frac{J_0}{2\sqrt{\pi}\mathcal{B}v}
\matt{1}{\sqrt{v-1}/\beta}{-\sqrt{v-1}/\beta}{(1+v)/\beta^2}.
\end{equation}
Here, we have defined $v\equiv 1+ (2-\beta\mu)^2$ and the dimensionless strength of the
magnetic field $\mathcal{B}\equiv |q|BR \sqrt{\beta}/(\sqrt{2m} c)$. 
The quantity $J_0\equiv (2\pi)^{\frac{3}{2}}\sqrt{m} R\exp[\beta\mu]
/\beta^{\frac{3}{2}}$ corresponds to the total particle current flowing into 
the scattering region at thermal equilibrium, i.e., for $\Delta T_i=
\Delta\mu_i=0$, as one can easily infer from (\ref{Jout}). 

The maximum efficiency in this strong field regime $\mathcal{B}\gg 1$
follows by inserting (\ref{ApproxOM}) into (\ref{MaxEffZT}) as
\begin{equation}\label{ZT_MaxEff_HF}
\eta_{{{\rm max}}}
= \eta_{{{\rm C}}}
\frac{\sqrt{2v}-\sqrt{1+v}}{\sqrt{2v}+\sqrt{1+v}}
\quad {{\rm with}} \quad
\mathcal{Z}T=  \frac{v-1}{2v}.
\end{equation}
The bounds (\ref{New_BoundZT}) and (\ref{New_BoundMaxEff}) are indeed
reached for $v\rightarrow\infty$, i.e., for $\beta\mu\rightarrow -\infty$
\footnote{ The limit $\beta\mu\rightarrow+\infty$ is incompatible with 
our classical approach, since it would lead to an exponentially high
equilibrium fugacity $\exp[\beta\mu]$ in the reservoirs \cite{Callen1985}.}.
However, in this limit, the equilibrium current $J_0 \sim \exp[\beta\mu]$, and likewise the Onsager matrix 
(\ref{ApproxOM}), decay exponentially.
Thus, the saturation of the bounds (\ref{New_BoundZT}) and
(\ref{New_BoundMaxEff}) comes at the price of vanishing power. 

\begin{figure}
\epsfig{file=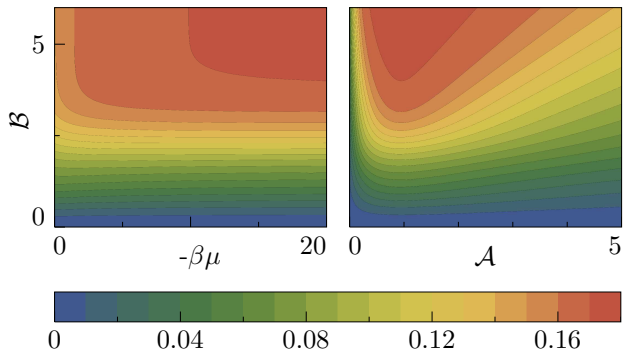, scale=.92}
\caption{Maximum efficiency. The left panel shows $\eta_{{{\rm max}}}/
\eta_{{{\rm C}}}$ as a function of the rescaled magnetic field 
$\mathcal{B}$ and $\beta\mu$ for aspect ratio $\ar=1$.
The right panel shows the dependence on $\mathcal{B}$ and $\ar$ for 
$\beta\mu=-20$.
\label{Fig_MaxEfficiency}}
\end{figure}

\emph{Efficiency diagrams}.--
Relaxing the assumption $\mathcal{B}\gg 1$, we now turn to an arbitrary 
field strength.
By repeating the procedure outlined in the 
preceding section using the full relation (\ref{Delta_s})
instead of the approximation (\ref{Delta_s_Approx}) as a starting point, 
we obtain closed, analytical expressions for the transmission coefficients 
$\mathcal{T}_{ji}(E)$ now depending explicitly on $\mathcal{B}$ and 
$\ar$ \cite{SM}.
After evaluating the primary Onsager coefficients (\ref{PrimOC}) 
numerically and including the boundary conditions (\ref{BoundaryCond}), 
we calculate the maximum efficiency $\eta_{{{\rm max}}}$, which is 
plotted in Fig. \ref{Fig_MaxEfficiency}.
We find that, for any $\mathcal{A}$ and $\beta\mu$, $\eta_{{{\rm max}}}$ 
vanishes at $\mathcal{B}=0$ as expected, since the vertical heat flux and 
the horizontal particle current decouple for vanishing magnetic field.
As $\mathcal{B}$ is increased, $\eta_{{\rm max}}$ grows monotonically.
Notably, if the aspect ratio deviates significantly from $1$,
larger values of $\mathcal{B}$ are necessary for $\eta_{{\rm max}}$ to
approach its upper bound (\ref{New_BoundMaxEff}). 
This effect is readily understood by recalling that for 
the high field scenario to apply, $\Delta s \sim 1/\mathcal{B}$ for a typical trajectory must be smaller than $\min\{l_1,l_2\} =\pi R \min\{1,\ar\}/(1+\ar)$ to ensure that only transitions between 
adjacent reservoirs are relevant. 
Our numerical results suggest an optimal aspect ratio $\ar^{\ast}(\mathcal{B},\beta \mu)$
close to $1$. 
Exploring this issue in more detail will be left to future work.

\emph{Efficiency at maximum power}.--
After studying the maximum efficiency of our device, we now consider
another important benchmark for the performance of a thermoelectric 
engine, its efficiency at maximum power $\eta^{\ast}$
\cite{Curzon1975, VandenBroeck2005, Esposito2009,Seifert2012},
which is obtained by maximizing the output power $P=-\Delta\mu_4 J_4^{\varrho}$ with respect to $\Delta\mu_4$. 
Expressed in terms of $\mathcal{Z}T$, this parameter reads 
\begin{equation}\label{Eff@MaxP}
\eta^{\ast}\equiv \eta_{{{\rm CA}}} \mathcal{Z}T/(2-\mathcal{Z}T),
\end{equation}
where $\eta_{{{\rm CA}}}=\eta_{{{\rm C}}}/2$ denotes the Curzon-Ahlborn 
value \cite{Curzon1975}, which is attained for $\mathcal{Z}T\rightarrow 1$. 
However, the constraint (\ref{New_BoundZT}) implies the stronger bound 
\begin{equation}\label{NewBound_Eff@MaxP}
\eta^{\ast}\leq \eta_{{{\rm CA}}}/3\simeq 0.167\eta_{{{\rm C}}}.
\end{equation} 
In the strong field regime, (\ref{Eff@MaxP}) becomes 
$\eta^{\ast}=\eta_{{{\rm CA}}}(v-1)/(3v+1)$.
Thus, like $\eta_{{\rm max}}$, $\eta^{\ast}$
reaches the bound (\ref{NewBound_Eff@MaxP}) only in the limit 
$v\rightarrow \infty$, i.e., for $\beta\mu\rightarrow-\infty$.
We can refrain from showing numerical data for $\eta^{\ast}$, 
since they are practically the same as those for $\eta_{{{\rm max}}}$.
Specifically, we have $0.97 <\eta^{\ast}/\eta_{{{\rm max}}} \leq 1$ throughout the whole parameter range of $\mathcal{B}$, $\ar$ and
$\beta\mu$ due to $\eta_{{{\rm max}}}-\eta^{\ast} = 
(\mathcal{Z}T)^3/64 + \mathcal{O}\left((\mathcal{Z}T)^4\right)$.

\emph{Concluding perspectives}.--
In this letter, we have introduced a classical formalism to describe 
heat and particle transport in non-interacting systems,
which can be regarded as the classical analogue to the well-established
Landauer-B\"uttiker approach. 
The crucial quantities of this formalism are the energy-dependent 
transmission coefficients $\mathcal{T}_{ji}(E)$, for which we 
have proven the sum rules (\ref{Sum_rules}). 
We emphasize that these sum rules follow solely 
from Liouville's theorem and thus hold for any kind of Hamiltonian 
dynamics inside a scattering region of arbitrary shape. 

For a Nernst geometry, 
in which a heat current is coupled to a 
perpendicular particle current via a magnetic field, a universal 
bound significantly lower than the Carnot value constrains the 
maximum efficiency. 
This bound can indeed be saturated for a strong field and 
exponentially small fugacities in the reservoirs.
The same bound holds for a cooling device based on the Ettings\-hausen effect \cite{Goldsmid2009},
\footnote{To operate the model as a refrigerator, we have to choose $\Delta \mu_4$ such that, 
for $\Delta T_3 < 0$, $J_3^q > 0$. 
The performance of the resulting device is benchmarked 
by the coefficient $\varepsilon\equiv J^q_3/(\Delta\mu_4 J^{\varrho}_4)$.
Maximizing $\varepsilon$ with respect to $\Delta\mu_4$ under the 
condition $J_3^q>0$ yields
\begin{equation}
\varepsilon_{{{\rm max}}}=\varepsilon_{{{\rm C}}} 
\frac{1-\sqrt{1-\mathcal{Z}T}}{1+\sqrt{1-\mathcal{Z}T}},
\end{equation}
where $\varepsilon_{{{\rm C}}}\equiv - T/\Delta T_3$ denotes the
coefficient of performance of an ideal refrigerator. 
As for the Nernst engine, the constraint (19) implies the bound 
$\varepsilon_{{{\rm max}}}/\varepsilon_{{{\rm C}}} \leq 3-2\sqrt{2}$,
which can be saturated only at the price of vanishing heat current
within the high field limit. }.
In both cases, this bound would not change even in the presence of an additional 
potential or for a geometrically deformed scattering region 
provided the two mirror symmetries are kept.
It remains an open question, however, whether these bounds 
would also apply if one included inelastic scattering or 
particle-particle interactions. 
Finally, due to its simplicity and physical transparency, our 
classical approach can provide a valuable benchmark for 
assessing the role of quantum effects in future modeling.

\begin{acknowledgments}
U.S. acknowledges support from ESF through the EPSD network. K.S. was supported by MEXT (23740289). 
\end{acknowledgments}

\end{document}



\title{Supplemental Material for A Classical Nernst Engine}


\author{}
\affiliation{}


\date{\today}

\begin{abstract}
\end{abstract}


\newcommand{\Jout}{J^{\varrho+}_i}
\newcommand{\Jin}{J^{\varrho -}_i}
\newcommand{\vectt}[2]{\left( \! \begin{array}{c}
#1\\ #2 \end{array} \! \right)}
\newcommand{\vecttt}[3]{\left[ \! \begin{array}{c}
#1 \\ #2\\ #3 \end{array}\! \right]}
\newcommand{\matt}[4]{\left( \! \begin{array}{cc}
#1 & #2\\ #3 & #4 \end{array} \! \right)}
\newcommand{\dd}[0]{d}
\newcommand{\IN}[0]{\text{in}}
\newcommand{\OUT}[0]{\text{out}}
\newcommand{\bs}[0]{\boldsymbol}

\newtheorem{lemma}{Lemma}
\newtheorem{corollary}{Corollary}
\newtheorem{definition}{Definition}
\newtheorem{proposition}{Proposition}
\newtheorem{theorem}{Theorem}

\setcounter{page}{6}
\section*{Supplemental Material for A Classical Nernst Engine}
\subsection*{Proof of the Sum Rules (7)} 
We will prove the sum rules
\begin{equation}\label{Sum_Rules}
\sum_i \mathcal{T}_{ji}(E) = 2 l_j
\quad {{\rm and}} \quad
\sum_j \mathcal{T}_{ji}(E) = 2 l_i
\end{equation}
for the transmission coefficients $\mathcal{T}_{ji}(E)$ introduced in 
Eq. (6) of the main text. To this end, we consider a particle with fixed 
energy $E$ injected at the position $s_{{{\rm in}}}$ and angle $\vartheta_{
{{\rm in}}}$.
After traveling through the circular scattering region the particle
eventually leaves it at the position 
$s_{{{\rm out}}}$ and angle $\vartheta_{{{\rm out}}}$. 
Since we assume Hamiltonian dynamics inside the scattering region, there is 
a one-to-one mapping 
\begin{equation}
\mathcal{M}: \begin{pmatrix} s_{\IN} \\ \vartheta_{\IN} \end{pmatrix} \mapsto
\begin{pmatrix} 
s_{\OUT} \\ \vartheta_{\OUT} \end{pmatrix} =
\begin{pmatrix} 
s_{\OUT}(s_{\IN}, \vartheta_{\IN}) \\ \vartheta_{\OUT}(s_{\IN}, \vartheta_{\IN}) \end{pmatrix}.
\end{equation}
The conditional probability $\tau_j(s_{\IN},\vartheta_{\IN})$, introduced in 
Eq. (5) of the main text, for a particle to reach the reservoir $C_j$ for fixed 
initial conditions $\vartheta_{{{\rm in}}}$ and $s_{{{\rm in}}}$ reads
\begin{equation}
\tau_j(s_{\IN}, \vartheta_{\IN}) = \int_{l_j} \! \dd s' \int_{-\pi/2}^{\pi/2} \! \dd \vartheta'  \, \delta(s' - s_{\OUT}) \delta (\vartheta' - \vartheta_{\OUT}).
\end{equation}
By recalling definition (6) of the main text, 
\begin{equation}
\mathcal{T}_{ji}(E) \equiv \int_{l_i} \! \dd s_{\IN} \int_{-\pi/2}^{\pi/2} \! \dd \vartheta_{\IN} \, \tau_j(s_{\IN}, \vartheta_{\IN}) \cos \vartheta_{\IN} \label{eq:transmission_coefficients},
\end{equation}
and using $\sum_j l_j = 2 \pi R$ we obtain the first of the sum rules 
(\ref{Sum_Rules}). To prove the second one, we change integration variables 
in \eqref{eq:transmission_coefficients} according to 
\begin{equation}\label{Coord_Change}
\begin{pmatrix} s_{\IN} \\ \vartheta_{\IN} \end{pmatrix} \rightarrow \begin{pmatrix} s_{\OUT} \\ \vartheta_{\OUT} \end{pmatrix}, 
\end{equation}
thus ending up with
\begin{equation}
  \sum_i \mathcal{T}_{ji}  = \int_{-\pi/2}^{\pi/2} \! \dd \vartheta' \int_{l_j} \! \dd s' \int_0^{2\pi R} \! \dd s_{\OUT} \int_{-\pi/2}^{\pi/2} \! \dd \vartheta_{\OUT} \, \delta( s' - s_{\OUT}) \delta(\vartheta' - \vartheta_{\OUT}) \cos \left[ \vartheta_{\IN}(s_{\OUT},\vartheta_{\OUT}) \right] \,  J(s_{\OUT},\vartheta_{\OUT}), \label{eq:sum_rule_functional_determinant}
 \end{equation}
where 
\begin{equation}\label{Jacobian}
 J(s_{\OUT},\vartheta_{\OUT}) \equiv \left| \frac{\partial (s_{\IN}, \vartheta_{\IN})}{\partial (s_{\OUT},\vartheta_{\OUT})} \right|
\end{equation}
denotes the Jacobian of 
the coordinate change (\ref{Coord_Change}), which can be determined as follows.
We introduce the coordinates $\bs{Q} \equiv (r, s)$, $\bs{P} \equiv (p_r, p_{s})$ 
in the four dimensional phase space $\Gamma$ of a single particle that are
connected to the Cartesian coordinates 
$ (x,y)$ and the corresponding momenta $(p_x, p_y)$ via the canonical 
transformation
 \begin{equation}
 x = r \cos \left[ s/R \right], \ \ y = r \sin \left[ s/R \right], \ \ p_r = m \dot{r}, \ \ p_{s} = m r^2 \dot{s}/R^2.
 \end{equation}
Next, within $\Gamma$, we define a two-dimensional surface $\mathcal{S}$ by
\begin{equation}
r = \text{const} \equiv R
\quad {{\rm and}} \quad H(\mathbf{Q},\mathbf{P})= \text{const} \equiv E,
\end{equation}
where $H(\mathbf{Q},\mathbf{P})$ denotes the single particle Hamiltonian.
This surface, which can be parametrized by the coordinates $(s, p_{s})$,
is pierced exactly twice by the phase space trajectory.
The two intersection points $(s_{\IN},p_{s_{\IN}})$ and
$(s_{\OUT},p_{s_{\OUT}})$ are connected by the one-to-one mapping
 \begin{equation}
  \mathcal{M}': \begin{pmatrix} s_{\IN} \\ p_{s_{\IN}} \end{pmatrix} \mapsto \begin{pmatrix} s_{\OUT}(s_{\IN}, p_{s_{\IN}} ) \\ p_{s_{\OUT}}(s_{\IN}, p_{s_{\IN}} ) \end{pmatrix}. 
 \end{equation}
From the Poincar\'e-Cartan theorem \cite{Ott2002}, it follows that the map
$\mathcal{M}'$ is volume conserving on $\mathcal{S}$, i.e., 
 \begin{equation}
  \left| \frac{\partial (s_{\IN}, p_{s_{\IN}})}{\partial (s_{\OUT}, p_{s_{\OUT}})} \right| = 1.
 \end{equation}
Rewriting (\ref{Jacobian}) as 
 \begin{equation}
  J(s_{\OUT},\vartheta_{\OUT}) =  \left| \frac{\partial (s_{\IN}, \vartheta_{\IN})}{\partial (s_{\IN}, p_{s_{\IN}})} \right| \left| \frac{\partial (s_{\IN}, p_{s_{\IN}})}{\partial (s_{\OUT}, p_{s_{\OUT}})} \right| \left| \frac{\partial (s_{\OUT}, p_{s_{\OUT}})}{\partial (s_{\OUT}, \vartheta_{\OUT})} \right| 
 \end{equation}
 leads to 
 \begin{equation}\label{Jacobian2}
  J(s_{\OUT},\vartheta_{\OUT}) = \left| \left( \frac{\partial p_{s_{\IN}}}{\partial \vartheta_{\IN}} \right)_{s_{\IN}} \right|^{-1} \left| \left( \frac{\partial p_{s_{\OUT}}}{\partial \vartheta_{\OUT}} \right)_{s_{\OUT}} \right|. 
 \end{equation}
Given the relations 
\begin{equation}
p_{s_{\IN}} = \sqrt{ 2 m E} \sin \vartheta_{\IN}
\quad {{{\rm and}}} \quad
p_{s_{\OUT}} = \sqrt{ 2 m E} \sin \vartheta_{\OUT},
\end{equation}
the partial derivatives showing up in (\ref{Jacobian2}) are readily calculated.
Thus, we finally arrive at
 \begin{equation}
  J(s_{\OUT},\vartheta_{\OUT}) = \frac{\cos \vartheta_{\OUT}}{\cos \left[ \vartheta_{\IN}(s_{\OUT}, \vartheta_{\OUT}) \right]}.
 \end{equation}
Inserting this result into equation \eqref{eq:sum_rule_functional_determinant} 
gives the second sum rule (\ref{Sum_Rules}).
 
We emphasize that our proof ultimately relies on the volume-preserving 
property of Hamiltonian dynamics and therefore applies irrespective of 
a potential inside the scattering region or the number of reservoirs 
attached to it. 
For simplicity, we have focused here on a circular scattering region.
However, generalization to arbitrary geometries is straight forward.

\subsection{Proof of Relation (18)}

We will prove that the Hermitian matrix 
\begin{equation}\label{K_Matrix}
\mathbb{K}\equiv \mathbb{L}+\mathbb{L}^t + i(\mathbb{L}-\mathbb{L}^t)
\end{equation}
introduced in Eq. (17) of the main text is positive semi-definite on 
$\mathbb{C}^2$. To this end, we need some mathematical prerequisites,
which we develop first. 

\begin{definition}\label{Def_AI}
Let $\mathbb{A}\in \mathbb{R}^{n\times n}$ be positive semi-definite, 
then the asymmetry index of $\mathbb{A}$ is defined as \cite{Brandner2013a}
\begin{equation}
\mathcal{S}(\mathbb{A})\equiv \min \left\{ \left. s\in\mathbb{R}
\right| \forall \mathbf{z}\in\mathbb{C}^n \quad 
\mathbf{z}^{\dagger} \left( s \left(\mathbb{A}+\mathbb{A}^t\right)
+i\left(\mathbb{A}-\mathbb{A}^t\right)\right)\mathbf{z}\geq 0 \right\}.
\end{equation}
\end{definition}

\begin{definition}
Let $\mathbb{A}\in\mathbb{R}^{n\times n}$ be a matrix with elements 
$A_{ij}\equiv (\mathbb{A})_{ij}\geq 0$, then $\mathbb{A}$ is 
sum-symmetric if \cite{Bapat2009}
\begin{equation}
\sum_i A_{ij} = \sum_i A_{ji}.
\end{equation}
\end{definition}

\begin{definition}\label{Def_Circuit_Matrix}
Let $\alpha \subseteq \{1,\dots,n\}$ be an index set 
and denote by $S(\alpha)$ the symmetric group on $\alpha$. 
For any permutation $\pi\in S(\alpha)$, the matrix $\mathbb{B}(\alpha)
\in \{0,1\}^{n\times n}$ with elements 
\begin{equation}
(\mathbb{B}(\alpha))_{ij}\equiv 
\begin{cases}
1 & {{\rm for}}\quad i,j \in \alpha \;\; {{\rm and}} \;\; j=\pi(i)\\
0 & {{\rm else}}
\end{cases}
\end{equation}
is a circuit matrix \cite{Bapat2009}.
\end{definition}

\begin{lemma}\label{Lem_AI_Circuit_Matrix}
For $\alpha$ like in Definition \ref{Def_Circuit_Matrix} let 
$\mathbb{E}(\alpha)\in\{0,1\}^{n\times n}$ be a matrix with 
elements $E_{ij}(\alpha)\equiv (\mathbb{E}(\alpha))_{ij}$ such that
$E_{ij}=1$ for $i=j\in\alpha$ and $E_{ij}=0$ otherwise.
Then, for any circuit matrix $\mathbb{B}(\alpha)$, the matrix 
$\mathbb{E}(\alpha)-\mathbb{B}(\alpha)$ is positive semi-definite
and its asymmetry index fulfills 
\begin{equation}
\mathcal{S}(\mathbb{E}(\alpha)-\mathbb{B}(\alpha)) \leq \cot
\left[ \frac{\pi}{|\alpha|} \right],
\end{equation}
where $|\alpha|$ denotes the cardinality of $\alpha$. 
\end{lemma}

\begin{proof}
Let $\mathbb{Q}\in\{0,1\}^{n\times n}$ be a permutation matrix such that $\mathbb{Q}^t\mathbb{E}(\alpha)\mathbb{Q}= \mathbbm{1}\oplus \mathbb{
O}_{n-|\alpha|}$, where $\mathbb{O}_m$ denotes the $m\times m$ matrix 
with only zero entries. 
Since the rows and columns of $\mathbb{B}(\alpha)$ 
indexed by $\alpha$ contain precisely one entry $1$, while the 
remaining rows and columns contain only zero entries, we have 
$\mathbb{Q}^t(\mathbb{E}(\alpha)-\mathbb{B}(\alpha))\mathbb{Q}
=(\mathbbm{1}-\mathbb{P})\oplus\mathbb{O}_{n-|\alpha|}$, where 
$\mathbb{P}\in\{0,1\}^{|\alpha|\times |\alpha|}$ is a permutation matrix.
The matrix $\mathbbm{1}-\mathbb{P}$ is positive semi-definite by virtue
of Theorem 1 of Ref. \cite{Brandner2013a}. Consequently, we have proven
the first assertion of Lemma \ref{Lem_AI_Circuit_Matrix}. Now, by recalling
Definition \ref{Def_AI}, we find 
\begin{equation}
\mathcal{S}\left(\mathbb{E}(\alpha)-\mathbb{B}(\alpha)\right)
=\mathcal{S}\left(\mathbb{Q}^t(\mathbb{E}(\alpha)-\mathbb{B}(\alpha))
\mathbb{Q}\right)
=\mathcal{S}\left((\mathbbm{1}-\mathbb{P})\oplus\mathbb{O}_{n-|\alpha|}
\right)
=\mathcal{S}\left(\mathbbm{1}-\mathbb{P}\right)
\leq \cot \left[\frac{\pi}{|\alpha|}\right],
\end{equation}
where the last inequality again follows from Theorem 1 of Ref. 
\cite{Brandner2013a}.
\end{proof}
\begin{corollary}\label{Cor_AI_SSMatrix}
Let $\mathbb{A}\in\mathbb{R}^{n\times n}$ be sum-symmetric with row sums 
$r_i\geq 0$ and $\mathbb{D}\in \mathbb{R}^{n\times n}$ a diagonal matrix 
with entries $(\mathbb{D})_{ij} \equiv r_i \delta_{ij}$. 
Then the matrix $\mathbb{D}-\mathbb{A}$ is positive semi-definite and its 
asymmetry index fulfills 
\begin{equation}
\mathcal{S}\left(\mathbb{D}-\mathbb{A}\right)\leq \cot 
\left[ \frac{\pi}{n} \right]. 
\end{equation}
\end{corollary}
\begin{proof}
It can be shown that for any sum-symmetric matrix $\mathbb{A}$ 
there is a finite set of circuit matrices $\mathbb{B}_k(\alpha_k)$ such that 
\begin{equation}
\mathbb{A}=\sum_k \lambda_k \mathbb{B}_k(\alpha_k),
\end{equation}
where the $\lambda_k$ are positive real numbers \cite{Bapat2009}.
Accordingly, the matrix $\mathbb{D}-\mathbb{A}$ can be rewritten as 
\begin{equation}
\mathbb{D}-\mathbb{A}=\sum_k \lambda_k\left( \mathbb{E}(\alpha_k)
-\mathbb{B}(\alpha_k)\right)
\end{equation}
with $\mathbb{E}(\alpha_k)$ like in Lemma \ref{Lem_AI_Circuit_Matrix}.
From Lemma \ref{Lem_AI_Circuit_Matrix}, it follows that 
$\mathbb{D}-\mathbb{A}$ is positive semi-definite. 
By using the convexity of the asymmetry index, proven in 
Proposition 2 of Ref. \cite{Brandner2013a}, and again Lemma
\ref{Lem_AI_Circuit_Matrix}, we infer 
\begin{equation}
\mathcal{S}\left(\mathbb{D}-\mathbb{A}\right)
\leq \max \left\{ \mathcal{S} \left( \mathbb{E}(
\alpha_k)- \mathbb{B}_k(\alpha_k)\right)\right\}
\leq \max\left\{ \cot \left[ \frac{\pi}{|\alpha_k|} \right] \right\}
\leq \cot \left[ \frac{\pi}{n}\right].
\end{equation}
\end{proof}

Coming back to our task of proving that the matrix (\ref{K_Matrix}) 
is positive semi-definite we consider the quadratic form
\begin{equation}\label{Quad_Form}
Q(\mathbf{z},s) \equiv \mathbf{z}^{\dagger} \left(
s\left(\mathbb{L}'+ {\mathbb{L}'}^t\right)+
i\left(\mathbb{L}'- {\mathbb{L}'}^t\right) \right) \mathbf{z}
\end{equation}
for arbitrary $\mathbf{z}\in\mathbb{C}^8$ and 
\begin{equation}\label{Lprime_Matrix}
\mathbb{L}'\equiv \left(\! \begin{array}{ccc}
\mathbb{L}'_{11}  & \cdots & \mathbb{L}'_{14}\\
\vdots            & \ddots & \vdots\\
\mathbb{L}'_{41}  & \cdots & \mathbb{L}'_{44}
\end{array} \! \right)
\equiv \int_0^{\infty} dE \; u(E) 
\left(\mathbb{D}-\mathbb{T}(E)\right)\otimes 
\left( \! \begin{array}{cc}
1      & E-\mu\\
E-\mu  & (E-\mu)^2
\end{array}\!\right)
\quad {{\rm with}} \quad
\mathbb{L}'_{ij}\equiv\left(\! \begin{array}{cc}
L_{ij}^{\varrho\varrho} & L_{ij}^{\varrho q}\\
L_{ij}^{q\varrho}       & L_{ij}^{qq}
\end{array} \! \right),
\end{equation}
where the $L_{ij}^{\kappa\nu}$ for $\kappa,\nu = \varrho, q$ denote the
primary Onsager coefficients defined in Eq. (12) of the main text. 
The matrix $\mathbb{T}(E)\in\mathbb{R}^{4\times 4}$ contains the 
transmission coefficients $\mathcal{T}_{ij}(E)$, i.e., $(\mathbb{T}(E))_{
ij}\equiv \mathcal{T}_{ij}(E)$, and $\mathbb{D}\in\mathbb{R}^{4\times 4}$ 
denotes a diagonal matrix with entries $(\mathbb{D})_{ij}=2l_i\delta_{ij}$.
By putting 
\begin{equation}
\mathbf{z}\equiv 
\mathbf{z}_1 \otimes \left(\!\begin{array}{c} 1 \\ 0 \end{array}\! \right)+
\mathbf{z}_2 \otimes \left(\!\begin{array}{c} 0 \\ 1 \end{array}\! \right)
\quad {{\rm with}} \quad
\mathbf{z}_1, \mathbf{z}_2\in\mathbb{C}^4
\end{equation}
and inserting this decomposition as well as (\ref{Lprime_Matrix}) into 
(\ref{Quad_Form}), we obtain 
\begin{equation}
Q(\mathbf{z},s)= \int_0^{\infty} dE \; u(E) \mathbf{y}^{\dagger}(E) 
\mathbb{K}'(s,E) \mathbf{y}(E),
\end{equation}
where $\mathbf{y}(E)\equiv \mathbf{z}_1 + (E-\mu)\mathbf{z}_2$. 
The Hermitian matrix $\mathbb{K}'(s,E)$ is given by 
\begin{equation}
\mathbb{K}'(s,E)\equiv 
s\left( (\mathbb{D} - \mathbb{T}(E))+ (\mathbb{D}-\mathbb{T}(E))^t \right)+
i\left( (\mathbb{D} - \mathbb{T}(E))- (\mathbb{D}-\mathbb{T}(E))^t \right)
\in \mathbb{C}^4.
\end{equation}
Comparing this expression with Definition \ref{Def_AI} and recognizing
that, thanks to the sum rules (\ref{Sum_Rules}), Corollary 
\ref{Cor_AI_SSMatrix} applies
to the matrix $\mathbb{D}-\mathbb{T}(E)$ shows that 
$\mathbb{K}'(s,E)$ is positive semi-definite for any $s\geq \cot[\pi/4]=1$ independent of $E$. 
Thus, since $u(E)=\sqrt{2mE}\exp[-\beta(E-\mu)]$ is non-negative for any $0\leq E
<\infty$, the quadratic form $Q(\mathbf{z},s)$ must be positive semi-definite
for $s\geq 1$ and comparing (\ref{Quad_Form}) with Definition \ref{Def_AI}
leads to 
\begin{equation}\label{AI_Lprime}
\mathcal{S}\left(\mathbb{L}'\right)\leq 1.
\end{equation}
We now construct the matrix $\mathbb{L}$ from $\mathbb{L}'$ in three 
successive steps. First, we extract the principal submatrix $\bar{\mathbb{L}}'$
by deleting the second and third row and column of $\mathbb{L}'$. 
Second, we simultaneously interchange the rows and columns of $\bar{\mathbb{L}}'$, 
thus transforming $\bar{\mathbb{L}}'$ to $\mathbb{P}\bar{\mathbb{L}}'
\mathbb{P}^t$, where
\begin{equation}
\mathbb{P} \equiv \left(\! \begin{array}{cccccc}
0 & 0 & 0 & 0 & 1 & 0\\
0 & 0 & 0 & 1 & 0 & 0\\
0 & 0 & 0 & 0 & 0 & 1\\
0 & 0 & 1 & 0 & 0 & 0\\
1 & 0 & 0 & 0 & 0 & 0\\
0 & 1 & 0 & 0 & 0 & 0
\end{array}\! \right).
\end{equation}
Third, we choose the partitioning 
\begin{align}
\mathbb{P}\bar{\mathbb{L}}'\mathbb{P}^t\equiv
\left(\! \begin{array}{cc}
\mathbb{M}_{11} & \mathbb{M}_{12}\\
\mathbb{M}_{21} & \mathbb{M}_{22}
\end{array}\!\right)
\quad {{\rm with}} \quad
\mathbb{M}_{11} & \equiv \left(\! \begin{array}{cc}
L^{\varrho\varrho}_{44} & L^{\varrho q}_{43}\\
L^{q\varrho}_{34}       & L^{qq}_{33}
\end{array} \! \right), \quad & 
\mathbb{M}_{12}  & \equiv \left(\! \begin{array}{cccc}
L^{\varrho q}_{44} & L^{\varrho\varrho}_{43} & L^{\varrho\varrho}_{41}
& L^{\varrho q}_{42}\\
L^{qq}_{34} & L^{q\varrho}_{33} & L^{q\varrho}_{31} & L^{qq}_{32}
\end{array}\! \right), \qquad\\
\mathbb{M}_{21}& \equiv \left(\! \begin{array}{cccc}
L^{q\varrho}_{44} & L^{\varrho\varrho}_{34} & L^{\varrho\varrho}_{14}
& L^{q \varrho}_{24}\\
L^{qq}_{43} & L^{\varrho q}_{33} & L^{\varrho q}_{13} & L^{qq}_{23}
\end{array}\! \right)^t, & \quad
\mathbb{M}_{22} & \equiv \left(\! \begin{array}{cccc}
L^{qq}_{44} & L^{q\varrho}_{43} & L^{q\varrho}_{41} & L^{qq}_{42}\\
L^{\varrho q}_{34} & L^{\varrho\varrho}_{33} & L^{\varrho\varrho}_{31}
& L^{\varrho q}_{32}\\
L^{\varrho q}_{14} & L^{\varrho\varrho}_{13} & L^{\varrho\varrho}_{11}
& L^{\varrho q}_{12}\\
L^{qq}_{24} & L^{q\varrho}_{23} & L^{q\varrho}_{21} & L^{qq}_{22}
\end{array}\! \right)
\end{align}
and take the Schur complement \cite{Zhang2005} of $\mathbb{P} 
\bar{\mathbb{L}}'\mathbb{P}^t$ with respect to $\mathbb{M}_{22}$, thus
ending up with
\begin{equation}
\mathbb{L}= \mathbb{P}\bar{\mathbb{L}}'\mathbb{P}^t / \mathbb{M}_{22}
\equiv \mathbb{M}_{11} - \mathbb{M}_{12}\mathbb{M}_{22}^{-1}\mathbb{M}_{21}.
\end{equation}
Finally, by using the Propositions 3 and 4 of Ref. \cite{Brandner2013a}
and the result (\ref{AI_Lprime}), we find
\begin{equation}
\mathcal{S}\left(\mathbb{L}\right)
=\mathcal{S}\left(\mathbb{P}\bar{\mathbb{L}}'\mathbb{P}^t/\mathbb{M}_{22}\right)
\leq \mathcal{S}\left( \mathbb{P}\bar{\mathbb{L}}'\mathbb{P}^t\right)
=\mathcal{S}\left(\bar{\mathbb{L}}'\right)
\leq \mathcal{S}\left(\mathbb{L}'\right)
\leq 1.
\end{equation}
Consequently, by virtue of Definition \ref{Def_AI}, the matrix 
(\ref{K_Matrix}) must be positive semi-definite.

\subsection{Calculation of the Full Transmission Coefficients (6)}
In this section, we will calculate the energy dependent transmission 
coefficients $\mathcal{T}_{ji}(E)$, defined in Eq. (6) of the main text,
for the classical Nernst engine. 
To this end, we consider a particle of energy $E$ injected from the 
reservoir $C_i$ with the angle $\vartheta$ at the boundary position 
$s_{{{\rm in}}}$. 
After traveling through the scattering region, the particle is 
absorbed in the reservoir $C_j$ at the position $s_{{{\rm out}}}$.
A direct geometrical analysis shows that the length $\Delta s \equiv
s_{{{\rm out}}}-s_{{{\rm in}}}$ of the boundary segment the particle 
passes by is given by 
\begin{equation}\label{TC_Deltas}
\Delta s(\vartheta) = 2R \begin{cases} 
{{\rm arccot}} \left[(1+\nu\sin\vartheta)/(\nu\cos\vartheta)\right]+\pi
& {{\rm for}}  \quad \nu>1, \; \sin\vartheta < -1/\nu\\
{{\rm arccot}} \left[(1+\nu\sin\vartheta)/(\nu\cos\vartheta)\right]
& {{\rm else}}
\end{cases},
\end{equation}
where
\begin{equation}
\nu\equiv R_c/R = \sqrt{E/\varepsilon}
\quad {{\rm with}} \quad
\varepsilon\equiv q^2B^2R^2/(2mc^2).
\end{equation}
For $\nu\leq 1$, $\Delta s (\vartheta)$ has a global maximum 
\begin{equation}
\Delta s^{\ast}\equiv 2R\; {{\rm arccot}}
\left[ \sqrt{1-\nu^2}/\nu\right] = 2 R \arcsin \nu \leq \pi R,
\end{equation}
with respect to $\vartheta$, which it assumes for $\sin\vartheta = -\nu$. 
For $\nu>1$, $\Delta s(\vartheta)$ captures the whole
interval $[0,2\pi R)$ as $\vartheta$ runs from $-\pi/2$ to $\pi/2$.
In both bases, we can solve (\ref{TC_Deltas}) for $\sin \vartheta$, thus 
obtaining
\begin{equation}
\sin\left[\vartheta_{\pm}(\Delta s)\right] = h_{\pm}(\Delta s)
\quad {{\rm for}} \quad \nu\leq 1 \qquad {{\rm and}} \qquad
\sin\left[\vartheta_+(\Delta s)\right] = h_+(\Delta s)
\quad {{\rm for}} \quad \nu >1
\end{equation}
with
\begin{equation}
h_{\pm}(x)\equiv (1/\nu) \left(-\sin^2 \left[x/(2R)\right]
\pm \cos\left[x/(2R)\right]
\sqrt{\nu^2-\sin^2\left[x/(2R)\right]}\; \right).
\end{equation}
Using this result, a reflection coefficient with rescaled argument defined as
$\bar{\mathcal{T}}_{ii}(\nu)\equiv\mathcal{T}_{ii}(E(\nu))=\mathcal{T}_{ii}
(\varepsilon\nu^2)$ can be written as 
\begin{equation}\label{TC_Integrals}
\bar{\mathcal{T}}_{ii}(\nu) =
\begin{cases}
\int_{0}^{l_i} d\sigma_i \int_{-\pi/2}^{\pi/2} d\vartheta \; \cos\vartheta
-\int_{0}^{\Delta s^{\ast}} d\sigma_i 
\int_{\vartheta_-(\sigma_i)}^{\vartheta_+(\sigma_i)}
d\vartheta \; \cos\vartheta & {{\rm for}}\quad \nu \leq \alpha_i\\[6pt]
\int_{0}^{l_i} d\sigma_i \int_{-\pi/2}^{\pi/2} d\vartheta \; \cos\vartheta
-\int_{0}^{l_i} d\sigma_i 
\int_{\vartheta_-(\sigma_i)}^{\vartheta_+(\sigma_i)}
d\vartheta \; \cos\vartheta & {{\rm for}}\quad \alpha_i < \nu \leq 1\\[6pt]
\int_0^{l_i} d\sigma_i \int_{\vartheta_+(2\pi R)}^{\vartheta_+(\sigma_i+2\pi R-l_i)}
d\vartheta \; \cos\vartheta 
+\int_0^{l_i} d\sigma_i \int_{\vartheta_+(\sigma_i)}^{\vartheta_+(0)}
d\vartheta \; \cos\vartheta
& {{\rm for}}\quad 1< \nu.
\end{cases}
\end{equation}
with $\alpha_i\equiv \sin [l_i/(2R)]$ and  $\sigma_i\equiv s_i-s_{{{\rm in}}}$ 
(see Fig. 1 of the main text for the definition of $s_i$).
The integrals showing up in (\ref{TC_Integrals}) can be solved by standard
techniques giving
\begin{equation}\label{TC_T0}
\bar{\mathcal{T}}_{ii}(\nu) = \begin{cases}
2l_i- \pi R\nu & {{\rm for}}\quad \nu \leq \alpha_i\\
2l_i - 2 H(\alpha_i,\nu) & {{\rm for}}\quad \alpha_i < \nu
\end{cases}
\end{equation}
with
\begin{equation}
H(\alpha,\nu)\equiv R\left((\alpha/\nu)\sqrt{\nu^2-\alpha^2}
+\nu \arcsin[ \alpha/\nu] \right).
\end{equation}
Analogously, we can express the $\bar{\mathcal{T}}_{i+1i}(\nu)$. 
Using $l_i+l_{i+1}=\pi R$, we obtain
\begin{equation}
\bar{\mathcal{T}}_{i+1i}(\nu) \equiv \mathcal{T}_{i+1i}(\varepsilon \nu^2) = \begin{cases}
\int_0^{\Delta s^{\ast}} d\sigma_i
\int_{\vartheta_-(\sigma_i)}^{\vartheta_+(\sigma_i)} d\vartheta
\; \cos\vartheta & {{\rm for}} \quad \nu\leq\alpha_i\\[6pt]
\int_0^{l_i} d\sigma_i
\int_{\vartheta_-(\sigma_i)}^{\vartheta_+(\sigma_i)} d\vartheta
\; \cos\vartheta & {{\rm for}} \quad \alpha_i < \nu\leq\alpha_{i+1}\\[6pt]
\int_0^{l_i} d\sigma_i
\int_{\vartheta_-(\sigma_i)}^{\vartheta_+(\sigma_i)} d\vartheta
\; \cos\vartheta -
\int_0^{\Delta s^{\ast}-l_{i+1}} d\sigma_i
\int_{\vartheta_-(\sigma_i+l_{i+1})}^{\vartheta_+(\sigma_i+l_{i+1})}
d\vartheta \; \cos\vartheta & {{\rm for}} \quad \alpha_{i+1}< \nu \leq 1\\[6pt]
\int_0^{l_i} d\sigma_i 
\int_{\vartheta_+(\sigma_i+l_{i+1})}^{\vartheta_+(\sigma_i)} 
d\vartheta \; \cos\vartheta & {{\rm for}}\quad 1< \nu
\end{cases}
\end{equation}
for $l_{i}\leq l_{i+1}$ and
\begin{equation}
\bar{\mathcal{T}}_{i+1i}(\nu) = 
\begin{cases}
\int_0^{\Delta s^{\ast}} d\sigma_i 
\int_{\vartheta_-(\sigma_i)}^{\vartheta_+(\sigma_i)}
d\vartheta \; \cos\vartheta  & {{\rm for}} \quad \nu \leq \alpha_{i+1}\\[6pt]
\int_0^{\Delta s^{\ast}} d\sigma_i 
\int_{\vartheta_-(\sigma_i)}^{\vartheta_+(\sigma_i)} 
d\vartheta \; \cos\vartheta  
-\int_0^{\Delta s^{\ast}-l_{i+1}} d\sigma_i 
\int_{\vartheta_-(\sigma_i+l_{i+1})}^{\vartheta_+(\sigma_i+l_{i+1})}
d\vartheta \; \cos\vartheta 
& {{\rm for}} \quad \alpha_{i+1} < \nu \leq \alpha_i\\[6pt]
\int_0^{l_i} d\sigma_i \int_{\vartheta_-(\sigma_i)}^{\vartheta_+(\sigma_i)}
d\vartheta \; \cos \vartheta
-\int_0^{\Delta s^{\ast}-l_{i+1}} d\sigma_i 
\int_{\vartheta_-(\sigma_i+l_{i+1})}^{\vartheta_+(\sigma_i+l_{i+1})}
d\vartheta \cos\vartheta & {{\rm for}}\quad \alpha_i<\nu\leq 1\\[6pt]
\int_0^{l_i} d\sigma_i \int_{\vartheta_+(\sigma_i+l_{i+1})}^{\vartheta_+(\sigma_i)}
d\vartheta \; \cos\vartheta & {{\rm for}}\quad 1<\nu
\end{cases}
\end{equation}
for  $l_i>l_{i+1}$. Evaluating the integrals yields 
\begin{equation}\label{TC_T1}
\bar{\mathcal{T}}_{i+1i}(\nu) = \begin{cases}
\pi R \nu & {{\rm for}} \quad \nu\leq\alpha_i^{{{\rm min}}}\\
2 H(\alpha_i^{{{\rm min}}},\nu) 
& {{\rm for}} \quad \alpha_i^{{{\rm min}}} < \nu\leq\alpha_{i}^{{{\rm max}}}\\
2 H(\alpha_i^{{{\rm min}}},\nu) + 2 H(\alpha_{i}^{{{\rm max}}},\nu) -\pi R \nu
& {{\rm for}} \quad \alpha_{i}^{{{\rm max}}}< \nu \leq 1\\
H(\alpha_i^{{{\rm min}}},\nu) + H(\alpha_{i}^{{{\rm max}}},\nu) - H(1,\nu)
+ (R/\nu)\sin [l_i/R] 
& {{\rm for}} \quad 1<\nu
\end{cases},
\end{equation}
where $\alpha_i^{{{\rm min}}}\equiv \min \{\alpha_i, \alpha_{i+1}\}$ and 
$\alpha_i^{{{\rm max}}}\equiv \max \{\alpha_i, \alpha_{i+1}\}$.
Finally, we calculate
\begin{align}
\bar{\mathcal{T}}_{i+2i}(\nu) \equiv \mathcal{T}_{i+2i}(\varepsilon \nu^2) & = \begin{cases}
0 & {{\rm for}}\quad \nu \leq \alpha_{i+1}\\[6pt]
\int_{0}^{\Delta s^{\ast}-l_{i+1}} d\sigma_i
\int_{\vartheta_-(\sigma_i+l_{i+1})}^{\vartheta_+(\sigma_i+l_{i+1})}
d\vartheta \; \cos\vartheta & {{\rm for}}\quad \alpha_{i+1}<\nu\leq 1\\[6pt]
\int_0^{l_i} d\sigma_i
\int^{\vartheta_+(\sigma_i+l_{i+1})}_{\vartheta_+(\sigma_i+\pi R)} 
d\vartheta \; \cos\vartheta
& {{\rm for}}\quad 1<\nu
\end{cases}\\
& =\begin{cases}
0 & {{\rm for}}\quad \nu \leq\alpha_{i+1}\\
\pi R\nu - 2H(\alpha_{i+1},\nu) & {{\rm for}}\quad \alpha_{i+1}<\nu\leq 1\\
2H(1,\nu)-2H(\alpha_{i+1},\nu) & {{\rm for}}\quad 1<\nu
\end{cases}\label{TC_T2}
\end{align}
and 
\begin{align}
\bar{\mathcal{T}}_{i+3i}(\nu) \equiv \mathcal{T}_{i+3i}(\varepsilon \nu^2) & = \begin{cases}
0 & {{\rm for}}\quad \nu \leq 1\\[6pt]
\int_0^{l_i} d\sigma_i 
\int^{\vartheta_+(\sigma_i + \pi R)}_{\vartheta_+(\sigma_i+ 2\pi R -l_i)}
d\vartheta \; \cos\vartheta & {{\rm for}}\quad 1<\nu
\end{cases}\\
& = \begin{cases}
0 & {{\rm for}}\quad \nu \leq 1\\
H(\alpha_{i+1}, \nu) + H(\alpha_i,\nu) -H(1,\nu)- (R/\nu)\sin[l_i/R]
& {{\rm for}}\quad 1 < \nu
\end{cases}.\label{TC_T3}
\end{align}
We note that the sum rules
\begin{equation}
\sum_{i} \bar{\mathcal{T}}_{ji}=\sum_{i} \mathcal{T}_{ji} = 2 l_j 
\qquad {{\rm and}}\qquad
\sum_{j} \bar{\mathcal{T}}_{ji}=\sum_{j} \mathcal{T}_{ji} = 2 l_i
\end{equation}
are easily verified for the coefficients (\ref{TC_T0}), (\ref{TC_T1}), 
(\ref{TC_T2}) and (\ref{TC_T3}).

%